\documentclass{elsarticle}
\usepackage{color}
\usepackage{graphicx}
\usepackage{amsmath,amssymb}

\newcommand{\hmu}{\hat{\mu}}

\def\lsim{\raise0.3ex\hbox{$<$\kern-0.75em\raise-1.1ex\hbox{$\sim$}}}
\def\gsim{\raise0.3ex\hbox{$>$\kern-0.75em\raise-1.1ex\hbox{$\sim$}}}

\journal{Nuclear Physics A}

\begin{document}

\begin{frontmatter}
\title{Probing the QCD phase diagram with fluctuations\tnoteref{t1}}
\tnotetext[t1]{Contribution to ``45 Years of Nuclear Theory at Stony Brook: A Tribute to Gerald E.\ Brown''}
\author{Bengt Friman}
\address{GSI Helmholtzzentrum f\"ur Schwerionenforschung,\\
  D-64291 Darmstadt, Germany}


\begin{abstract} 
The relevance of higher order cumulants of conserved charges 
for the analysis of freeze-out and critical conditions
in heavy ion collisions at LHC and RHIC is discussed. Using properties of $O(4)$
scaling functions, the generic structure of these higher cumulants
at vanishing baryon chemical potential is discussed. Chiral model calculations
are then used to study their properties at non-zero baryon chemical potential.
It is argued that the rapid variation of sixth and higher 
order cumulants at the phase boundary may be used to explore the QCD phase diagram
in experiment. Moreover, results for the Polyakov loop susceptibilities in SU(3) lattice gauge theory
as well as in (2+1) flavor lattice QCD are discussed. An analysis of the ratios of susceptibilities  
indicates that the deconfinement transition is reflected in characteristic modifications of these ratios.

\end{abstract}
\end{frontmatter}

\section{Introduction}

Strongly interacting matter at high temperature or large net baryon
number density is expected to undergo a transition from a phase
where chiral symmetry is broken spontaneously to one where
chiral symmetry is restored. At vanishing baryon chemical
potential ($\mu_B=0$), this transition is of the crossover type for non-zero quark masses. At
$\mu_B > 0$ a second order phase transition point, the so-called
chiral critical end point, may exist. A large experimental
as well as theoretical effort is invested into the exploration of the QCD phase
diagram and the development of appropriate tools and observables
that can provide clear-cut signals for the existence of phase transitions
and their universal properties. In this context, fluctuations~\cite{koch}, 
in particular fluctuations of net charges~\cite{ste,raj},
may prove useful.

Critical behavior is connected with long range correlations and
enhanced fluctuations, due to the appearance
of soft modes in the neighborhood of a phase transition.
Fluctuations of baryon number and electric charge
have been shown to reflect the critical behavior of the chiral transition
\cite{lattice}.
In the exploration of the QCD phase diagram at non-zero temperature
and baryon chemical potential, higher order cumulants of the net baryon number
play a particularly important role, since they diverge on
the chiral phase transition line in the chiral limit (vanishing light quark masses, $m_{q}=0$) and at
the conjectured chiral critical end point.

Here I will  discuss robust features of cumulants
of net baryon number fluctuations that can be extracted from
considerations based on $O(4)$ universality, on existing lattice
calculations and on model calculations~\cite{Friman:2011pf}. 

It has recently been emphasized that also the deconfinement 
transition can be accurately characterized by exploring fluctuations~\cite{prd}. In this case the 
relevant fluctuations are those of the Polyakov loop, the order parameter 
for the spontaneous breaking of the $\mathcal{Z}(N)$ center symmetry in pure gauge theory.
I briefly discuss how this may be used to analyze the deconfinement 
transition~\cite{prd-2}.

\section{Charge fluctuations and $O(4)$ scaling functions}

In studies of critical fluctuations in nucleus-nucleus collisions, the hadron resonance gas (HRG) model, serves as a baseline, since it 
describes bulk observables~\cite{HRG,karsch,bazavov,Borsanyi:2011zz} and does not exhibit critical behavior. 
For sufficiently small quark masses, the fluctuations close to the
pseudo critical transition line $T_{pc}(\mu_B)$ are expected to 
reflect the universal properties \cite{Allton}
of the 3-dimensional, $O(4)$ symmetric, spin model \cite{Engels}.
For  fluctuations of the net baryon number,
the $O(4)$ scaling properties of high order cumulants
differ qualitatively from the predictions of the HRG model.
Lattice calculations of cumulants of the net baryon number and electric charge,
performed in the transition region at vanishing baryon chemical potential
and non-zero quark mass, are
indeed consistent with the $O(4)$ universality class~\cite{lattice}.

Moreover, lattice studies of (2+1) flavor QCD~\cite{Ejiri} show that for physical light quark masses, 
the magnetic equation of state is in the $O(N)$ scaling regime. Thus, we can expect that 
the critical fluctuations at the chiral transition are reflected in suitably chosen observables,
which may then be used to pin down the crossover transition in strongly interacting matter.
As I will argue, this is indeed the case for higher cumulants of net charges~\cite{Friman:2011pf}.

In nucleus-nucleus collisions, the 
particle multiplicities are well described in a thermal model using
the partition function of a hadron resonance gas \cite{HRG}. The analysis of
data obtained within the HRG model at high beam energies, corresponding to small values
of the baryon chemical potential, suggest that the freeze-out curve $T_f(\mu_B)$ 
is close to the expected QCD phase boundary. Indeed, at $\mu_B=0$
the chemical freeze-out seems to occur at or very close to the QCD crossover transition~\cite{Andronic:2011yq}.
Thus, it is possible that the critical fluctuations of the chiral transition are reflected in fluctuations of, e.g., net charges.

At $\mu_B/T \simeq 0$, the higher cumulants of net charges
can be computed within lattice QCD~\cite{lattice, bazavov}. Eventually such calculations will
provide a systematic theoretical framework for unravelling  
the relation of the freeze-out conditions at RHIC and LHC energies and the 
pseudo-critical line in the QCD phase diagram.
However, at present lattice calculations provide only limited information on sixth and higher 
order cumulants. Viable alternatives for discussing qualitative
features of the net baryon number fluctuations is offered by O(4) scaling theory and by
chiral effective models. In particular, effective models have the advantage
that they can be extended to $\mu_B > 0$ with a tractable effort. On the other hand,
a clear disadvantage of such models is that they do not account for the potentially
important contribution from resonances in the hadronic phase.

Close to the chiral critical point, at $T=T_c, \mu_{q}=0, m_{q}=0$, the free energy density may be represented in terms of a
singular and a regular contribution\footnote{From now on I use the chemical potential for quarks $\mu_{q}$ rather than that 
for baryon number, $\mu_{B}=3\mu_{q}$.}
\begin{equation}
f(T,\mu_q,m_q) = f_s(T,\mu_q,m_q) + f_r(T,\mu_q,m_q).
\label{freeenergy}
\end{equation}
Higher order derivatives of the free
energy density with respect to temperature or chemical potential
are dominated by the non-analytic (singular) part,
$f_s$.  The explicit dependence of the free energy on the chemical potentials of
electric charge and strangeness as well as  on the
strange quark mass is suppressed for simplicity.
The singular part of the free energy may be written in the scaling form
\begin{equation}
\frac{f_s(T,\mu_q,h)}{T^4} =
A\, h^{1+1/\delta} f_f(z) \ ,\ z \equiv t/h^{1/\beta\delta},
\label{singular}
\end{equation}
where $\beta$ and $\delta$ are critical exponents of the 3-dimensional
$O(4)$ spin model~\cite{Engels}, $A$ is a constant, and the reduced temperature and the symmetry 
breaking external field are given by~\cite{Friman:2011pf}
\begin{eqnarray}
t &\equiv& \frac{1}{t_0}\left( \frac{T-T_c}{T_c} +
\kappa_q \left( \frac{\mu_q}{T_{c}}\right)^2
\right),
\nonumber \\
h &\equiv& \frac{1}{h_0} \frac{m_q}{T_c}.
\label{scalingfields}
\end{eqnarray}
Here $T_c$ is the critical temperature in the chiral limit
while $t_0$ and $h_0$ are non-universal scale parameters \cite{Ejiri}.
The dependence of the reduced temperature $t$ on 
the quark chemical potential $\mu_{q}$, with $\kappa_{q}\simeq 0.06$, is fixed by charge conjugation symmetry and 
by requiring that the curvature of the chiral phase boundary at small $\mu_{q}/T$ is reproduced~\cite{Mukherjee}.
 
The scaling properties of the order parameter $M$ are characterized by  
$M = h^{1/\delta} f_G(z)$, with the scaling function
\begin{equation}
f_G(z) = -\left( 1+\frac{1}{\delta} \right) f_f(z) + \frac{z}{\beta\delta}
f'_f(z).
\label{fG}
\end{equation}
The scaling function $f_f(z)$ and its derivatives $f_f^{(n)}(z)$ are known
for $n\le 3$  \cite{Engels_2011}. These results can be utilized to 
determine the generic structure of higher order cumulants of
the net baryon number fluctuations.

The cumulants of net baryon number are obtained by differentiating the free energy density (\ref{freeenergy})
with respect to $\hmu_q = \mu_q/T$,
\begin{equation}
\chi_{n}^{B} = - \frac{1}{3^{n}}
\frac{\partial^{n}(f/T^4)}{\partial\hmu_q^n}.
\label{obs}
\end{equation}
The singular contribution to the cumulants at $\mu_{q}=0$ is then given by\footnote{For $\mu_{q}=0$, 
the odd cumulants vanish, owing to the quadratic dependence on $\mu_{q}$ in (\ref{scalingfields}).}
\begin{equation}
(\chi_n^B)_{s} \sim
- h^{(2-\alpha -n/2)/\beta\delta}\, f_f^{(n/2)}(z),
\label{fluct_mass}
\end{equation}
where I have introduced the specific-heat critical exponent $\alpha$, using the scaling relation $2-\alpha = \beta\delta (1+1/\delta)$.

Because $\alpha$ is negative for $O(4)$ spin systems in 3 dimensions
($\alpha = -0.2131$ \cite{Engels}), all cumulants 
of the net baryon number of order $n\leq 4$, 
remain finite at the second order transition in the chiral limit ($t=0,\, h\to 0$) for $\mu_q=0$. Thus,  the first divergent cumulant 
is obtained for $n= 6$. At non-zero $\mu_{q}$, the odd derivatives survive and the $n=3$ cumulant is the first one to diverge 
in the chiral limit. However, for small $\mu_{q}/T$, the singular part is suppressed by a factor $(\mu_{q}/T)^{3}$ relative to that of the 
sixth order cumulant~\cite{Friman:2011pf}.

\begin{figure}
\begin{center}
\includegraphics*[width=6.cm]{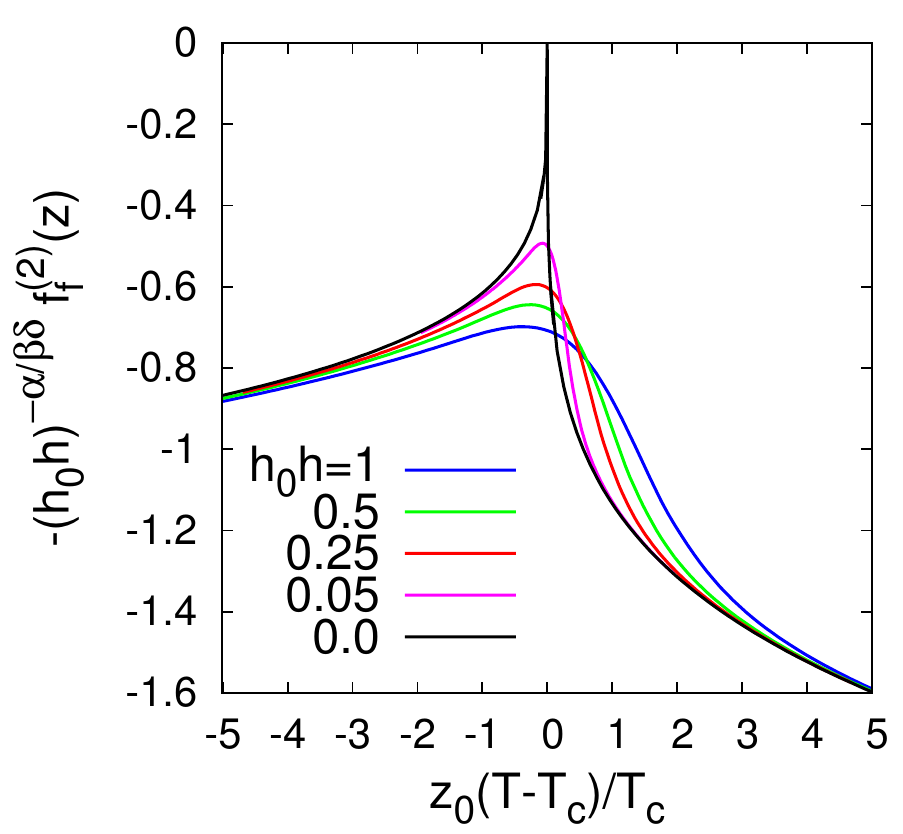}
\includegraphics*[width=6.cm]{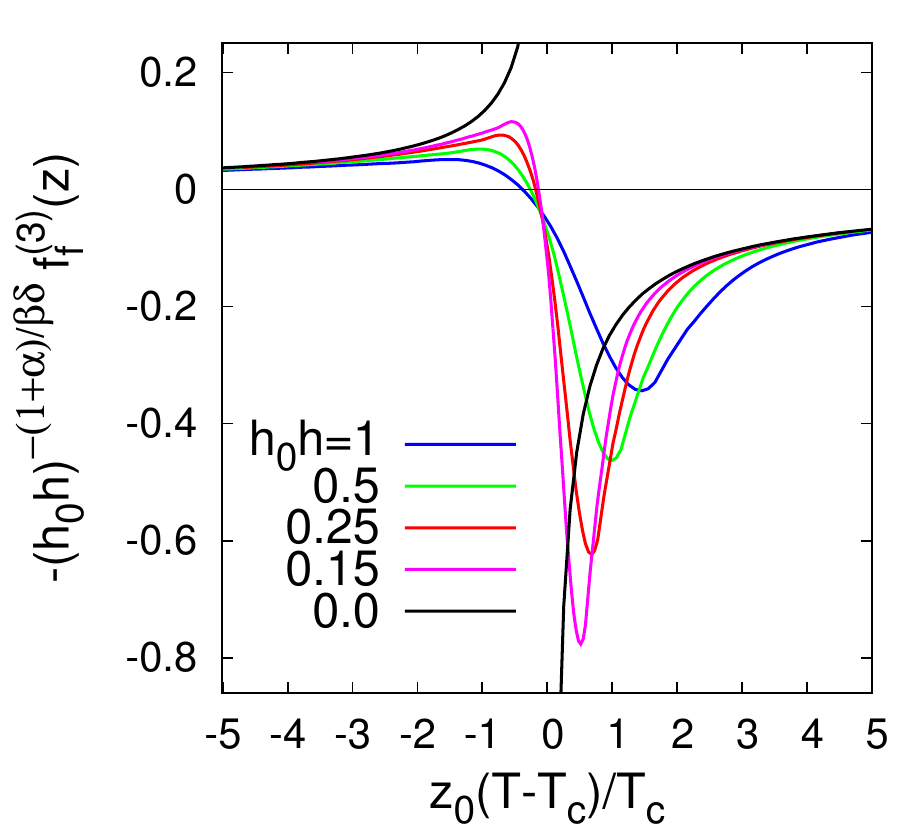}
\caption{
The non-analytic contributions to
$\chi_4^B$ (left) and $\chi_6^B$ (right) arising from
second and third derivatives of the singular part of the free energy. 
Here $h_0$ and $z_0$ 
are non-universal scale parameters \cite {Friman:2011pf}.
}
\label{fig:generic}
\end{center}
\end{figure}

Using Eq.~(\ref{fluct_mass}), one finds the leading
singularity in the chiral limit,
\begin{equation}
(\chi_n^B)_{s} \sim
- |t|^{2-\alpha -n/2} f^{(n/2)}_\pm,
\label{fluct}
\end{equation}
where
\begin{equation}
f^{(n)}_\pm = \lim_{z\rightarrow \pm \infty} |z|^{-(2-\alpha -n)} f_f^{(n)}(z).
\label{limit}
\end{equation}
Hence, the singular part of $\chi_4^B$, which is proportional to
$f_f^{(2)}$, has
the same structure as that of the specific heat; it is proportional to
the second derivative of the free energy with respect to temperature.
Universality arguments imply~\cite{Friman:2011pf} that in the chiral limit, the cumulants $\chi_{n}^{B}(t)$ of order $n\geq4$
are all positive for $t<0$ and alternate in sign for $t>0$.   
Consequently, at non-zero quark
mass, $h>0$, one expects $\chi_6^B$ to change sign in the transition
region and $\chi_8^B$ to do so twice, etc.. 
For a given $h>0$, this is reflected 
in the $z$-dependence of the scaling functions $f_f^{(n)}(z)$ of the 3-dimensional $O(4)$ model \cite{Engels_2011}, 
as shown in Fig.~\ref{fig:generic} for different values of the symmetry breaking
parameter $h$. Thus, the generic structure of the $n^{\rm th}$ order cumulant is determined by the corresponding $O(4)$ scaling function, 
in the chiral limit as well as for non-zero values of the quark mass. 

In the chiral limit, the non-analytic contribution to $\chi_4^B$ vanishes 
at the chiral transition temperature, $t=0$. Consequently, in the transition region,
the regular terms dominate in $\chi_4^B$ .
Nonetheless, the non-analytic term in $\chi_4^B$ varies rapidly with
temperature, leading to a pronounced maximum in the transition region,
observed in lattice as well as in model calculations.

The temperature at which $\chi_6^B$ changes sign is non-universal since it depends on the 
magnitude of the regular terms. However, in the scaling regime the location of the extrema and
the corresponding amplitudes follow universal scaling laws, as discussed 
in \cite{Friman:2011pf,Engels,Engels_2011}. Thus, Eq.~(\ref{fluct_mass}) implies that the 
divergent cumulants scale with the correlation length $\xi\sim h^{-\nu/\beta\delta}$ as
$\chi^{B}_{n}\sim \xi^{-(2-\alpha-n/2)/\nu}$. With the $O(4)$ critical exponents \cite{Engels}, $\alpha=-.2131$ 
and $\nu=.7377$, one finds $\chi^{B}_{6}\sim -\xi^{1.1}$ and $\chi^{B}_{8}\sim -\xi^{2.4}$ at $t=0$. Thus, as expected, the 
strength of the singularity grows with the order of the cumulant.

We now focus on the properties of the sixth order cumulant,
$\chi^B_{6}(T)$, or correspondingly on the ratio of cumulants
$R_{6,2}^B(T) = \chi^B_{6}(T)/\chi^B_{2}(T)$. Based on the characteristics of the scaling function, 
shown in Fig.~\ref{fig:generic}, it is clear that for a sufficiently small, but non-zero, quark mass, $\chi^B_{6}(T)$ has a
maximum in the hadronic phase, close to the transition region and then drops rapidly. Furthermore, just above the critical temperature 
in the chiral limit, $\chi^B_{6}(T)$ is negative and 
exhibits a sharp minimum. Lattice calculations of $\chi^B_{6}(T)$ \cite{Allton,Schmidt}
indicate that in QCD with physical quark masses, these 
basic features, which stem from the singular part of
$\chi^B_6$, persist. In particular, for the discussion below, it is important that
$\chi^B_{6}(T)< 0$ in the vicinity of the pseudo-critical temperature
for chiral symmetry restoration.

\begin{figure}
\vspace*{-0.8cm}
\begin{center}
\includegraphics*[width=6.5cm]{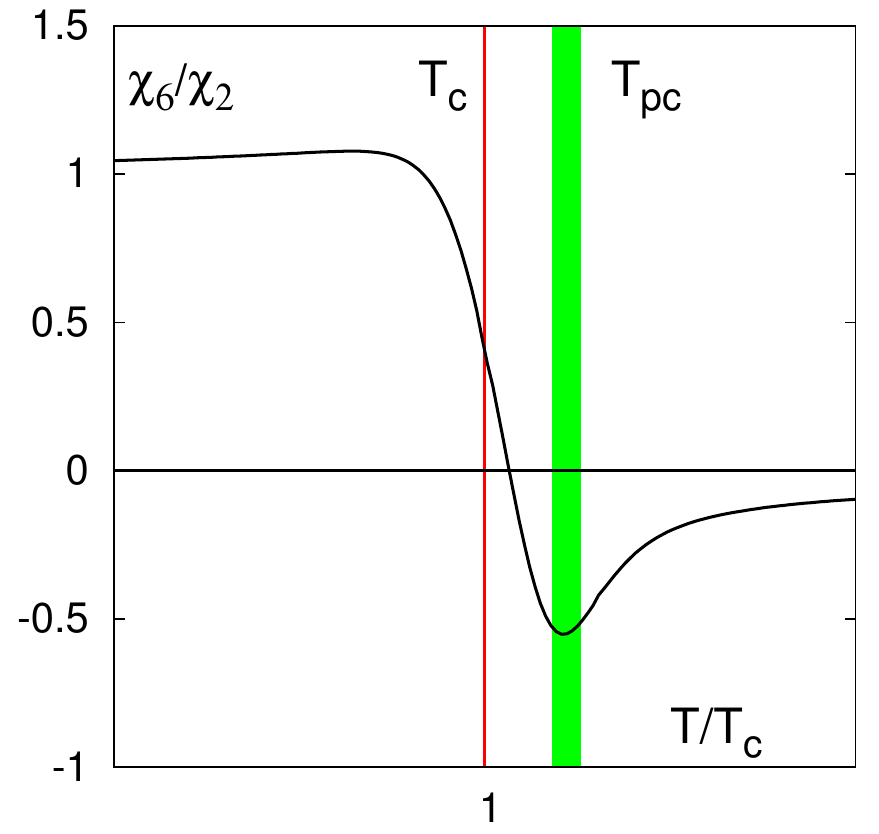}
\vspace*{-0.6cm}
\caption{Schematic plot of the ratio
of the sixth and second order cumulants of the net baryon number,
where $T_c$ is the critical temperature in the chiral limit, while 
the green band indicates the crossover transition in QCD with physical light quark masses.
The pseudo-critical temperature $T_{pc}$ corresponds to a peak in the chiral susceptibility.
}
\label{fig:schematic}
\end{center}
\end{figure}

The rapid temperature dependence makes the sixth order
cumulant a potential probe of the chiral crossover transition 
in heavy ion collisions. Indeed, if the
freeze-out occurs in a temperature range close to the chiral crossover
temperature, as indicated by the HRG analysis of particle multiplicities~\cite{Andronic:2011yq}, one would expect  the
data to show a negative sixth order cumulant, in striking contrast to the HRG result,
$R_{6,2}^B = 1$. In Fig.~\ref{fig:schematic},  
$R_{6,2}^B$ is shown in a schematic plot, where both the critical temperature in the chiral limit ($T_c$) and the crossover
temperature for chiral symmetry restoration for physical light quark masses ($T_{pc}$) are indicated. According to scaling arguments,
$T_{pc}>T_{c}$, as indicated in the figure. 

Higher order cumulants, e.g. $\chi^B_8$, also show a characteristic behavior near the chiral transition, 
owing to the singular part of the free energy~\cite{Friman:2011pf}. As noted above, the strength of the singularity, 
and thus the signature of critical fluctuations, grows with the order of the cumulant. Therefore, one may naively imagine 
that the prospects for unraveling the QCD chiral transition could be improved by considering higher cumulants. 
However, this advantage is counterbalanced  by the problems, which arise because with increasing order the 
cumulants are more and more sensitive to the tail of the probability distribution~\cite{morita2012net}.   

\section{Fluctuations at \boldmath $\mu_B/T > 0$}

In order to explore whether the rapid temperature dependence of $\chi_{6}^{B}$ can indeed be utilized 
to track the phase boundary, one must extend the discussion to 
non-vanishing baryon chemical potential. This is done in a chiral model, the
Polyakov loop extended quark meson model (PQM), within the framework of the functional
renormalization group \cite{Friman:2011pf,Skokov:2010uh}. The chiral transition of the PQM model 
belongs to the universality class of the 3-dimensional $O(4)$ spin system, while deconfinement
is modeled by a $\mathcal{Z}(3)$ symmetric effective potential for the Polyakov loop. Thus, 
in the light quark mass limit, this model yields the universal scaling functions, 
discussed in the previous section. Moreover, within this approach 
the higher cumulants can be computed as functions of $\mu_q$ and $T$. 
Details on the calculation can be found in~\cite{Friman:2011pf,Skokov:2010uh}.

\begin{figure}[t]
\begin{center}
\hspace*{-0.5cm}
\includegraphics*[width=6.cm,]{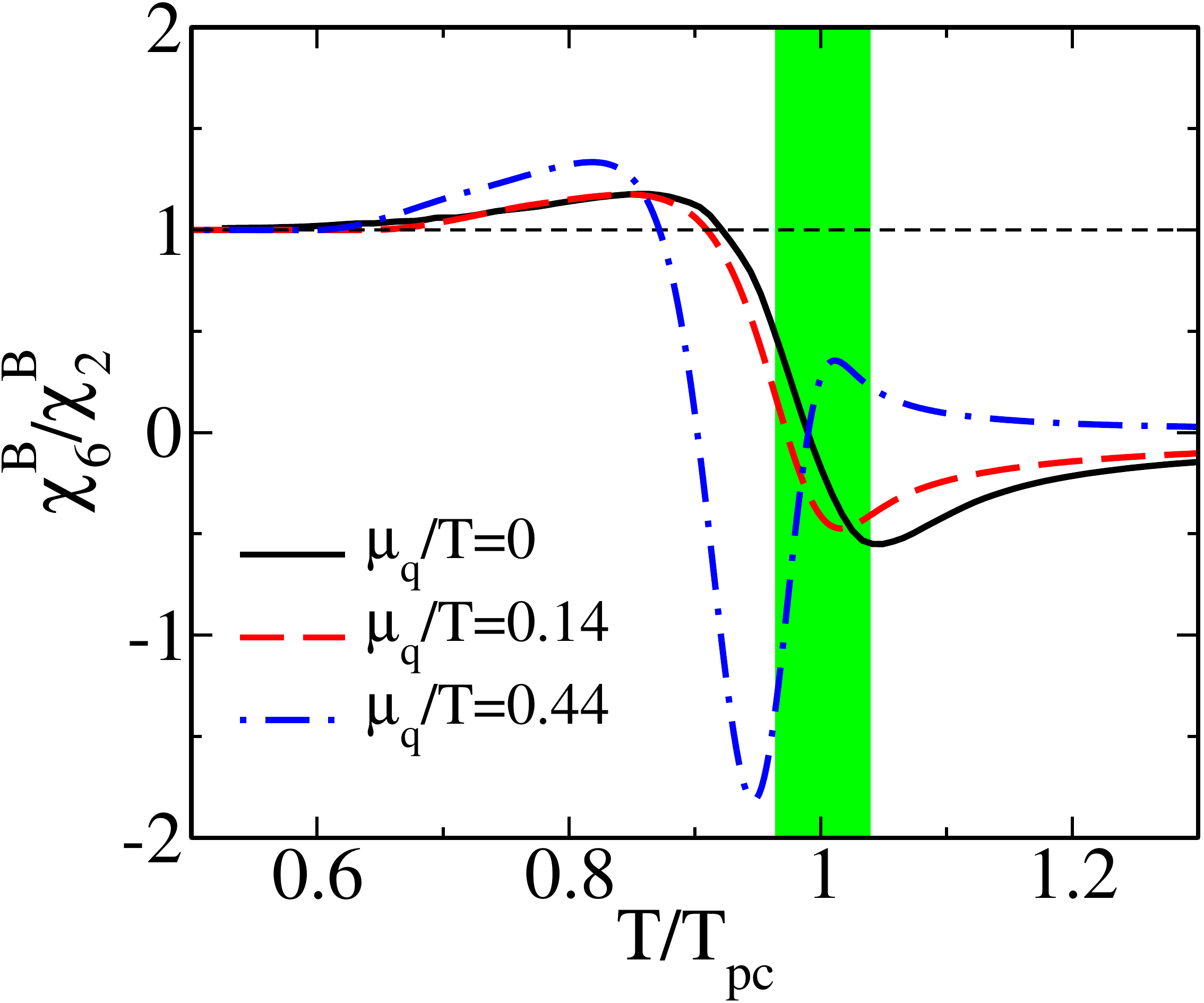}\hspace*{0.3cm}
\includegraphics*[width=6.cm,]{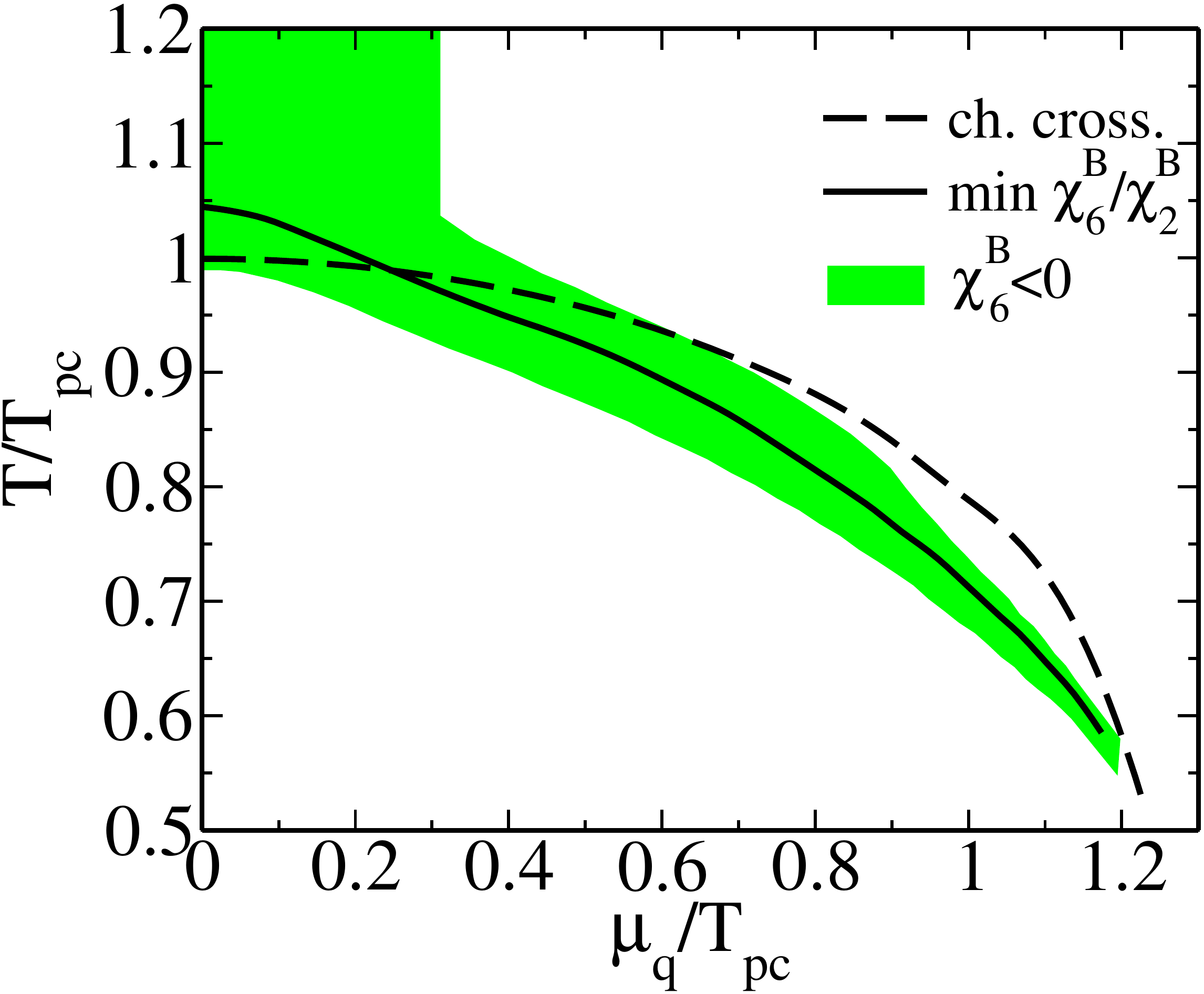}
\vskip 0.0cm\caption
{
On the left, the temperature dependence of the ratio $\chi_{6}^B$/$\chi_{2}^B$ 
for various $\mu_q/T$ is shown, while on the right, the location of the chiral crossover line [dashed] and the minimum
in $\chi_6^B$ [solid line] are displayed. The band shows the range where $\chi_6^B$ is negative. 
}
\label{fig:PD}
\end{center}
\end{figure}

The resulting ratio $R_{6,2}^B$ of the $6^{\rm th}$ and $2^{\rm nd}$ order cumulants
is shown for $\mu_q/T=0$ and $\mu_q/T > 0$ in the left panel of Fig.~\ref{fig:PD}. The
ratio approaches the hadron resonance gas result at low temperatures, and reproduces
the expected  $O(4)$ scaling properties in the transition region. In particular, a pronounced
minimum, with $R_{6,2}^B<0$, is obtained in the vicinity of the chiral crossover
temperature. Although the exact location of the minimum and its depth are
model dependent, the qualitative structure of the ratio is robust. 
With increasing $\mu_{q}/T$, the singular structure becomes stronger and the minimum 
is shifted to smaller temperatures. This dependence on $\mu_{q}/T$  can be understood
in terms of a Taylor expansion of $R_{6,2}^B$ about $\mu_{q}/T=0$, where 
the leading correction in $\mu_q/T$ is due to the (more singular) eighth order cumulant~\cite{Friman:2011pf}.

The line of minima of $R_{6,2}^B$
in the $\mu_q$-$T$ plane and the region where the ratio is negative are
shown in the right panel of Fig.~\ref{fig:PD}, together with the pseudo critical temperature
of the chiral crossover transition obtained with a physical pion mass.
At non-zero baryon chemical potential, the temperature interval where 
$\chi_6^B$ (and $R^B_{6,2}(\mu_q/T)$) is negative, shrinks and
closely follows the crossover transition line. 
The region where the sixth order cumulant $\chi_6^B$ is
negative, extends into the symmetry broken phase. Thus, 
if freeze-out occurs close to the chiral crossover temperature, the
sixth order cumulant of the net baryon number fluctuations will be negative
at LHC energies as well as for high RHIC beam energies.

The basic features discussed here for
net baryon number fluctuations also carry over to 
electric charge fluctuations, as indicated by lattice and 
model calculations~\cite{Cheng,Fu:2009wy,Skokov:2011rq}. Thus,  the corresponding ratio of electric charge cumulants also probes 
the conditions at freeze-out and their relation to the critical behavior in strongly interacting matter.
Moreover, the higher cumulants of electric charge are more directly accessible in experiment than those of the net baryon number.

It should be stressed that there are several uncertainties that may affect the data that are not accounted for in the results presented here.
These include acceptance corrections~\cite{Bzdak:2012ab,Kitazawa:2012at}, finite volume
effects~\cite{Bzdak:2012an}, volume fluctuations~\cite{Skokov:2012ds}  as well as possible 
non-equilibrium effects.

Preliminary data by the STAR collaboration~\cite{Chen:2013zaa} show a clear suppression of the ratio of the sixth order to second order 
net-proton number cumulant relative to the HRG result. The suppression depends strongly on the beam energy, with the strongest suppression
at the lower energies, while the dependence on the centrality of the collision is weak. At low beam energies, the STAR collaboration finds that also 
the ratio $\chi_{4}/\chi_{2}$ is suppressed relative to the hadron resonance gas ~\cite{Adamczyk:2013dal}. Such a suppression is expected, 
since at non-zero $\mu_{q}$, $\chi_{4}$ picks up a contribution from $\chi_{6}$, which is negative close to the phase boundary~\cite{Skokov:2012kw}. 
In view of the ideas presented  above, the STAR data are quite intriguing. However, a quantitative interpretation of the energy and centrality 
dependence of the cumulants is not yet available.


\section{Fluctuations at the deconfinement transition}

The deconfinement transition of QCD is connected with spontaneous breaking of the global $\mathcal{Z}(3)$ center symmetry
of color $SU(3)$. The symmetry is exact in the limit of infinitely heavy quarks and is explicitly broken by dynamical quark degrees of freedom.
In an $SU(3)$ Yang-Mills theory, the deconfinement transition is first order, while for physical light quark masses it is of the crossover type.

The Polyakov loop is an order parameter of confinement, which is linked with the free energy of a static quark immersed in a hot gluonic 
medium \cite{McLerran:1980pk,Kaczmarek:2002mc}.
At low temperatures, the thermal expectation value of the Polyakov loop   $\langle | L|\rangle $ vanishes,  indicating color confinement,
while at high temperatures  $\langle | L|\rangle\neq  0$, corresponding to a finite energy of a static quark, and consequently deconfinement of 
color and the spontaneous breaking of the $\mathcal{Z}(3)$ symmetry.

In the crossover regime, the order parameter varies smoothly, and does not provide an accurate signature for the transition temperature. 
Moreover, in Yang-Mills theory on a lattice the Polyakov loop exhibits finite size effects, which smoothens the first order transition, and thus complicates the 
determination of the transition temperature. It has been shown that the Polyakov loop susceptibilities, in particular their ratios, offer a useful probe
of deconfinement, also in full QCD with dynamical light quarks~\cite{prd,prd-2}.

For the color gauge groups $SU(N_c\ge 3)$, the Polyakov loop is complex-valued. Consequently, the fluctuations of the Polyakov loop 
are represented by susceptibilities along a longitudinal and a transverse direction\footnote{In the real sector of the Polyakov loop, longitudinal 
and transverse components correspond to the real and imaginary direction respectively. Although the thermal average of the imaginary part 
of the Polyakov loop vanishes in the real sector, the fluctuations in the imaginary direction do not.}. In lattice calculations, the absolute value 
of the Polyakov loop and the corresponding susceptibility are commonly used.

\section{ Polyakov loop susceptibilities on the lattice}

The temperature dependence of the Polyakov loop
susceptibilities was computed~\cite{prd,prd-2} within  SU(3) lattice gauge theory, using
the Symanzik improved gauge action on  $N_\sigma^3\times
N_\tau$ lattices for different   values of the temporal lattice sizes $N_\tau=(4,6,8)$   and for
spatial extensions $N_\sigma$ varying from 16 to 64. 

\begin{figure}[t]
 \center{\includegraphics[width=8.5cm]{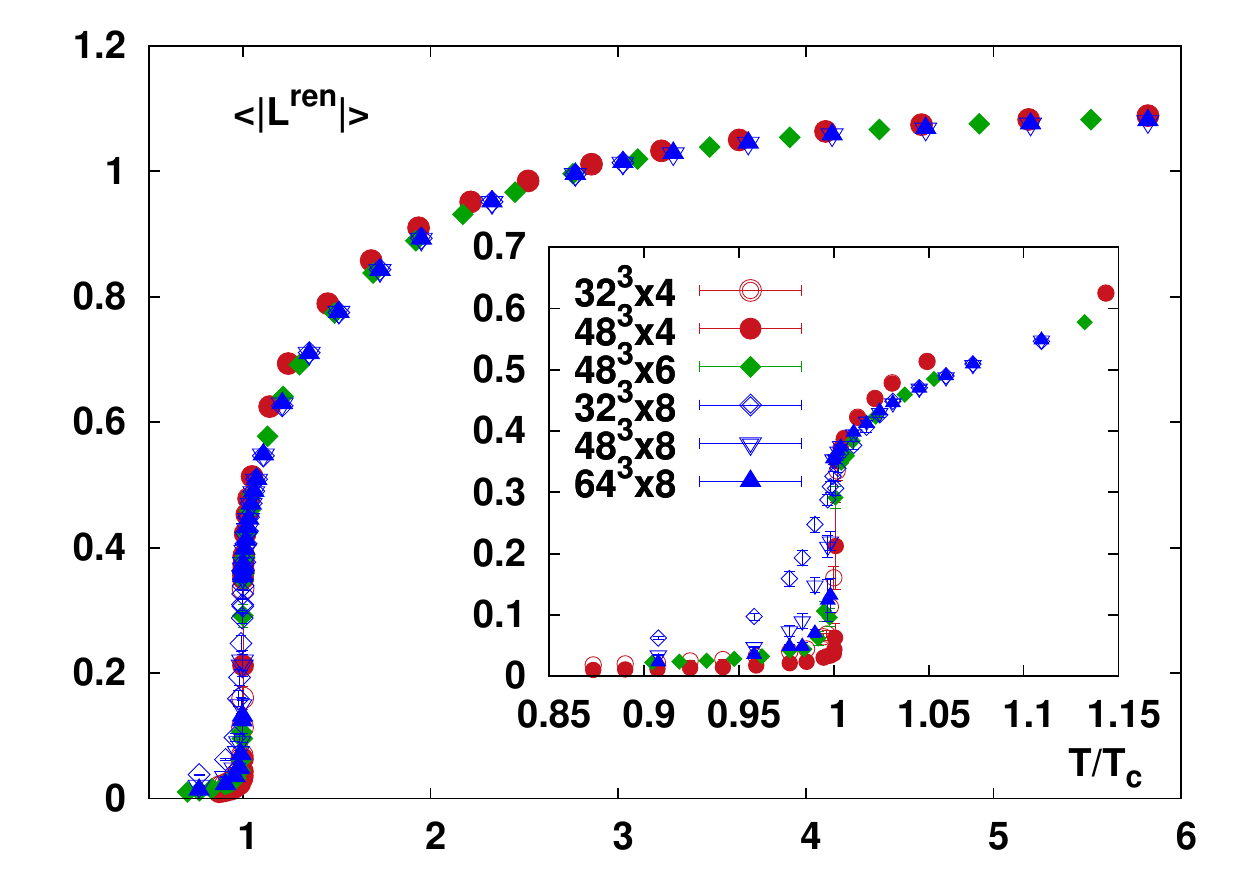}}
 \caption{The temperature dependence of  the modulus of the renormalized Polyakov loop  in  SU(3) gauge  theory, on various sized lattices.
 In the insert, finite size effects are clearly visible.
 } 
  \label{fig:polyakov}
\end{figure}

On a $N_\sigma^3\times N_\tau$ lattice,  the Polyakov loop is defined as the trace of the product over temporal gauge links,
\begin{equation}
L^{\rm bare} = \frac{1}{N_c N_\sigma^3}\sum_{\vec x}Tr \prod_{\tau=1}^{N_\tau} U_{(\vec x,\tau),4}
\end{equation}
while  the renormalized
Polyakov loop \cite{Kaczmarek:2002mc} is given by
\begin{eqnarray}\label{r1}
 L^{\rm ren} = \left(Z(g^2)\right)^{N_\tau} L^{\rm bare}.
\end{eqnarray} 
The ensemble average of the modulus of the Polyakov loop, $ \langle \vert L^{\rm ren} \vert \rangle$, is shown as a function of temperature in 
Fig.~\ref{fig:polyakov} for different sized lattices. This quantity is well defined in the continuum and thermodynamic limits. 
While no volume effects are visible in the deconfined
phase, the results at fixed $N_\tau$ in the confined phase, show the expected volume dependence, $\sim1/\sqrt V$.

Using the renormalized Polyakov loop, one can define the renormalized Polyakov loop
susceptibility
\begin{eqnarray}\label{eq:chiA}
T^3 \chi_A =& \frac{N_\sigma^3}{N_\tau^3} \left( \langle \vert L^{\rm ren} \vert^2 \rangle - \langle
   \vert L^{\rm ren} \vert \rangle^2\right).
\end{eqnarray}
As noted above, in color SU(3) the Polyakov loop operator is complex. Thus, in addition to $\chi_A$, it is useful to consider also
the longitudinal and transverse fluctuations of the Polyakov loop \cite{prd}

\begin{eqnarray}
T^3 \chi_{L} =& \frac{N_\sigma^3}{N_\tau^3} \left[ \langle  (L^{\rm ren}_{L})^2 \rangle - \langle
    L^{\rm ren}_{L} \rangle^2\right].\label{eq:chiL}\\
T^3 \chi_{T} =& \frac{N_\sigma^3}{N_\tau^3} \left[ \langle  (L^{\rm ren}_{T})^2 \rangle - \langle
    L^{\rm ren}_{T} \rangle^2\right],\label{eq:chiT}
    \end{eqnarray}

\begin{figure*}[!t]
 \includegraphics[width=6cm]{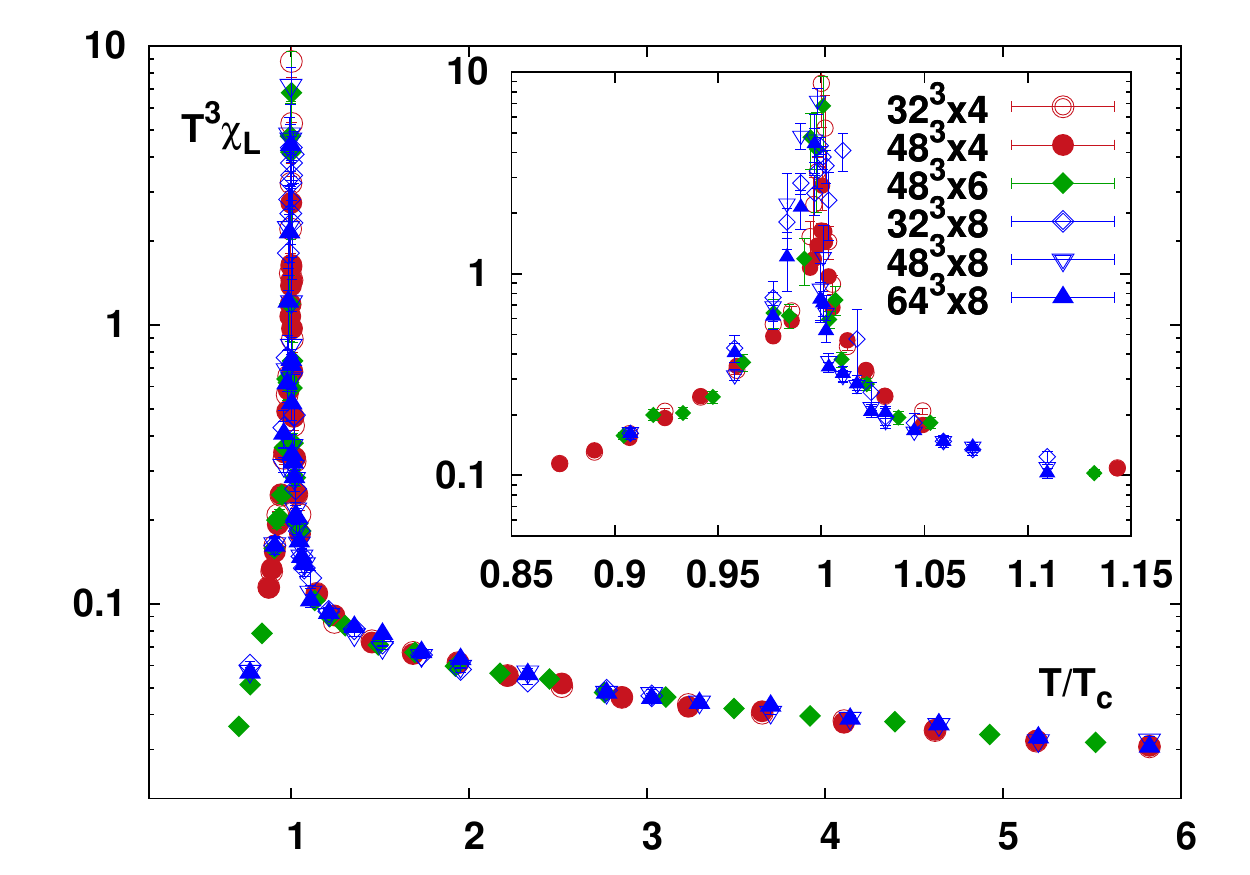}
 \includegraphics[width=6cm]{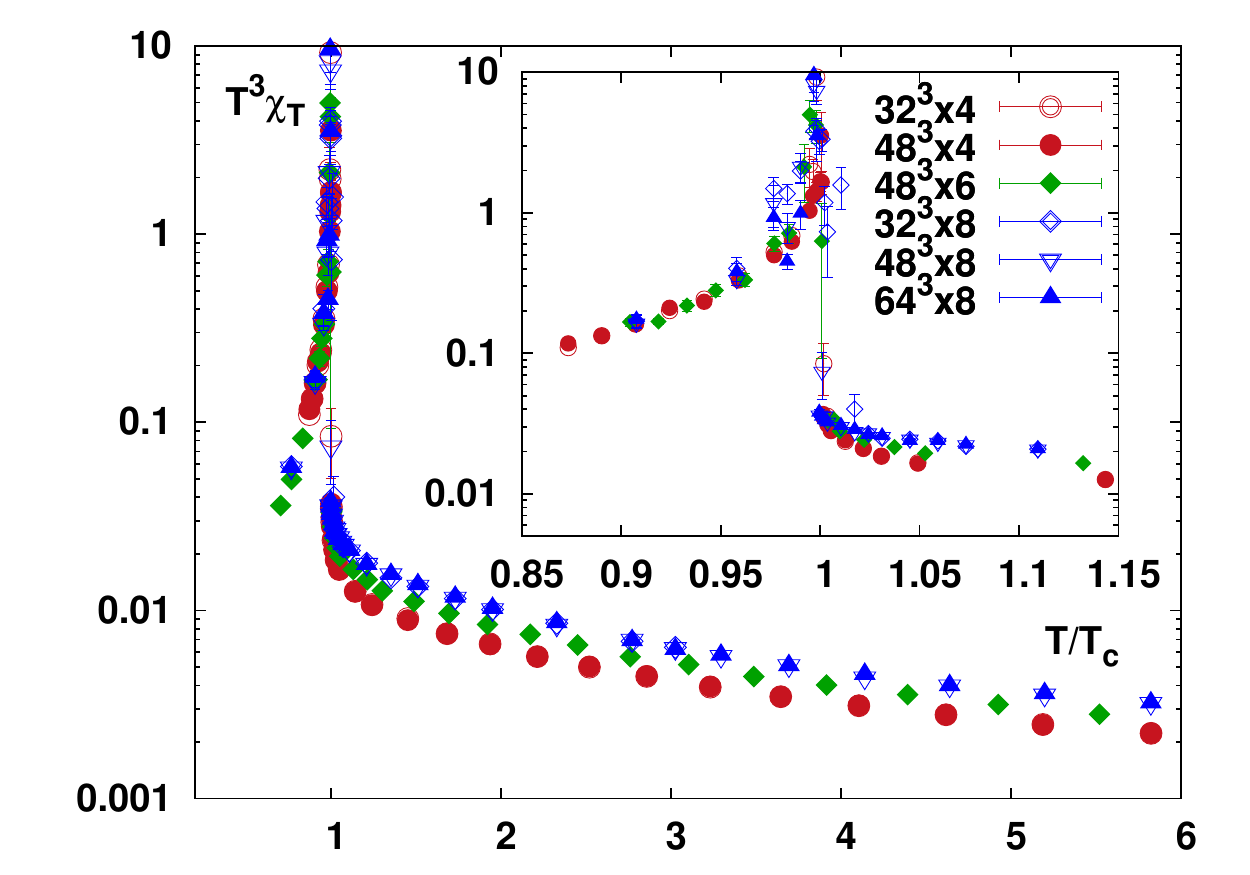}
  \caption{ The temperature dependence of the renormalized Polyakov loop susceptibilities from   Eqs. \eqref{eq:chiL} and \eqref{eq:chiT} on various lattice sizes, in SU(3) pure gauge theory.  The temperature is normalized to the critical value. }
 \label{fig:sus}
\end{figure*}

The longitudinal and transverse renormalized Polyakov loop susceptibilities obtained on different lattice
sizes are shown in Fig.~\ref{fig:sus}. Near the phase
transition, $0.95 < T/T_c < 1.05$, both susceptibilities show a rather strong dependence on the volume, 
consistent with the first order nature of the phase transition
in pure gauge theory. Outside this region, the longitudinal fluctuations of the Polyakov loop,
show only a minimal dependence on $N_\tau$ and $N_\sigma$ in both phases. This is the case also 
for the susceptibility of the modulus, $\chi_{A}$~\cite{prd}. The transverse susceptibility $\chi_{T}$,  however, exhibits a residual
$N_\tau$ dependence in the deconfined phase.

\begin{figure}[!ht]
 \includegraphics[width=6cm]{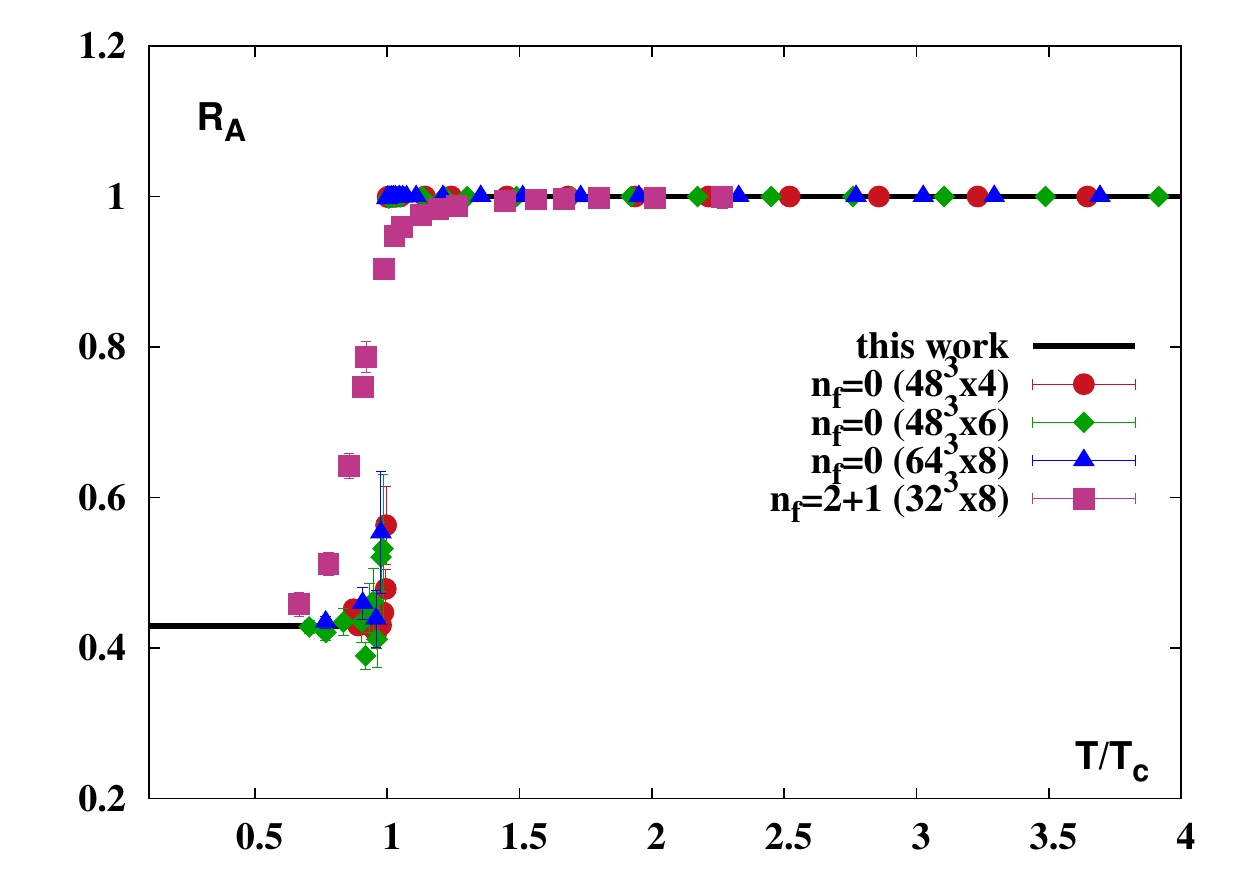}
\includegraphics[width=6cm]{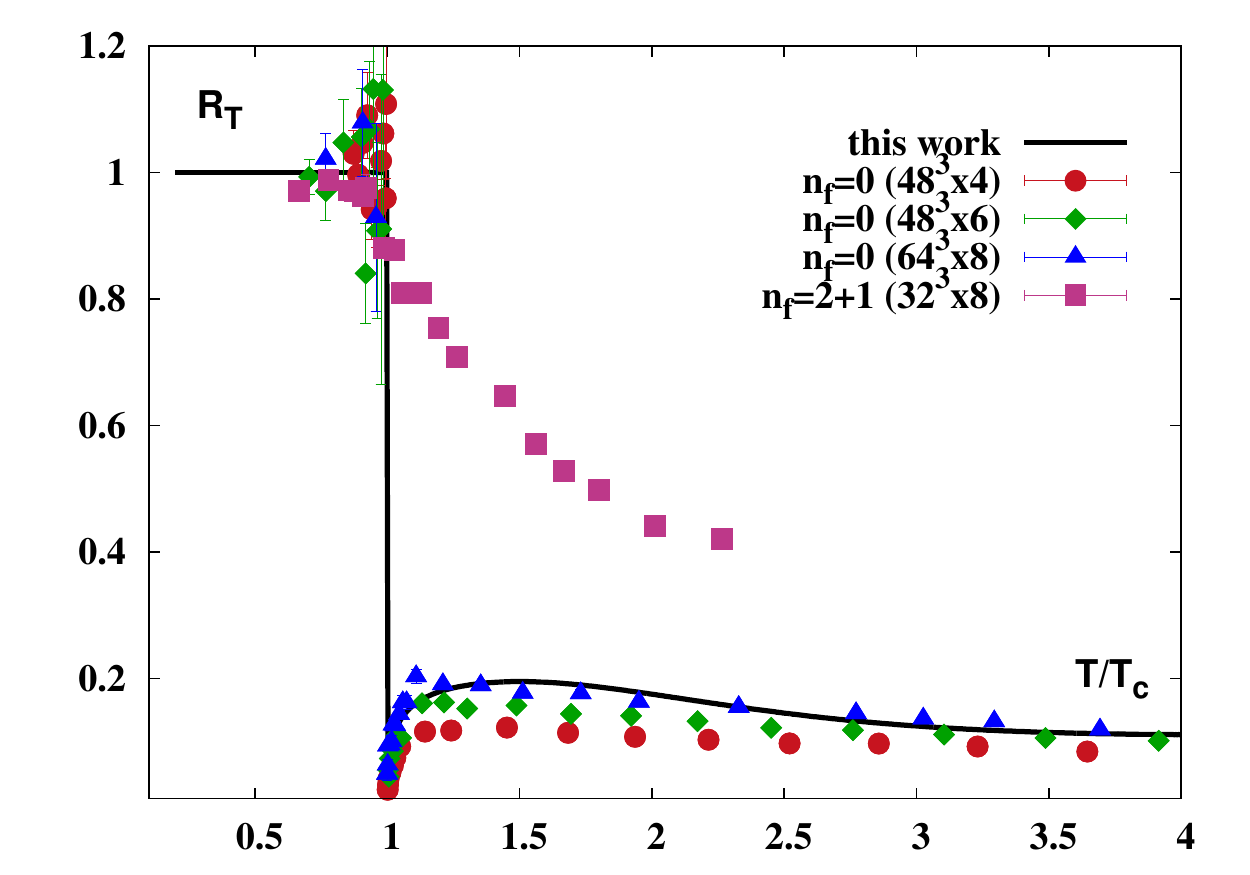}
 \caption{The ratios of Polyakov loop susceptibilities $R_{A}$ and $R_{T}$ obtained in lattice gauge theory for a pure gauge system 
 and for (2+1) flavor QCD. The temperature is normalized to its (pseudo) critical value for respective lattice. The lines show the model results, 
 as explained in the text.
 }
\label{ratioA}
\end{figure}

The ambiguities of the renormalization scheme can, to a large extent, be avoided by considering ratios of Polyakov loop susceptibilities~\cite{prd}.
In Fig.~\ref{ratioA} the ratios $R_{A}=\chi_{A}/\chi_{L}$ and $R_{T}=\chi_{T}/\chi_{L}$ of susceptibilities obtained in pure SU(3) gauge theory  are shown.
Both ratios are volume independent and exhibit a strong discontinuity at the deconfinement phase transition. 
A clear-cut interpretation of the almost temperature independent values of $R_{A}$ and $R_{T}$ in the confined phase is obtained 
using general considerations and the  $\mathcal{Z}(3)$ symmetry~\cite{prd}. Thus, e.g.,  for temperatures well below $T_{c}$  
the ratio $R_{A}\simeq 2-\pi/2$ indicates that the dominant contribution to the susceptibilities is due to Gaussian fluctuations. 

The properties of $R_T$ above $T_c$, shown 
in the right panel of Fig.~\ref{ratioA}, indicate that in the SU(3) pure gauge theory, the fluctuations  of the longitudinal part of the Polyakov loop are much stronger than those of
the transverse part. This is consistent with the shape of the effective Polyakov loop potential, which is ingrained primarily by the $SU(3)$ Haar measure~\cite{Sasaki:2006ww}.

Also shown in Fig.~\ref{ratioA} are results obtained by the HotQCD collaboration with a (2+1) flavor HISQ action and for almost physical quark masses~\cite{ejiri5,hisq}. For $N_f\neq 0$, the temperature  in Fig. \ref{ratioA}  is normalized to the corresponding pseudo critical temperature.
In the presence of dynamical quarks, the Polyakov loop is no longer an order parameter and remains non-zero even in the low temperature phase.  Consequently, the ratios of the Polyakov loop susceptibilities are modified due to  explicit breaking of the $\mathcal{Z}(3)$ symmetry. One therefore expects the smoothening of these ratios across the pseudo critical temperature. Indeed Fig. \ref{ratioA}  shows that, in the presence of dynamical quarks, both ratios vary continuously with temperature. The ratio $R_A$ interpolates between the two limiting values set by pure gauge theory, while the width of the crossover region varies with the number of flavors and the quark  masses.

In the deconfined phase, the ratio $R_T$ is strongly influenced by dynamical quarks. In addition to the stronger smoothening effect observed, this ratio deviates substantially 
from the pure gauge result at high $T$. However, also for small quark masses, the slopes of $R_T$ and $R_A$ change in a narrow temperature range
close to the transition point. A quantitative investigation of the effect of quarks on the Polyakov loop susceptibilities and their ratios at the deconfinement transition
require systematic studies of the dependence on system size and quark masses.

The lattice results have been used to construct an effective potential for the Polyakov loop, which is consistent with 
the mean value and the fluctuations of the Polyakov loop as well as with the bulk thermodynamics of pure $SU(3)$ gauge theory~\cite{prd-2}. 
The lines shown in Fig.~\ref{ratioA} show the model results for $R_{A}$ and $R_{T}$. 

\section{Concluding remarks}

I have discussed two aspects of fluctuations at the QCD phase boundary. On the one hand, lattice results and model calculations indicate that 
chiral critical fluctuations are reflected in the higher cumulants of net charges. First data on nucleus-nucleus collisions at RHIC show a suppression of
net proton cumulants, in qualitative agreement with theoretical expectations. However, further studies are needed in order to obtain a quantitative 
understanding of these phenomena. The deconfinement transition, on the other hand, is associated with fluctuations of the Polyakov loop. I
discussed how the Polyakov loop susceptibilities and, in particular, ratios thereof provide a useful probe of the deconfinement transition in
lattice QCD.  Also here further work is needed to obtain a quantitative probe of deconfinement.
                                                                                                                    
In the mid-70s, when I was a beginning graduate student in Finland, the two potential supervisors were abroad on sabbatical, one of them in Stony Brook.
Gerry then agreed to step in as my advisor. For some time I visited him regularly in Copenhagen. However, before long Gerry found that the arrangement 
was inefficient and suggested that I come to Stony Brook for my PhD. I am deeply grateful to him for his guidance, for many inspiring discussions on 
physics and on other topics and for his continuous support and friendly advise over many years. I dedicate this paper to his memory.

\section*{ Acknowledgments}
I would like to thank the organizers of this memorable meeting, Dima Kharze\-ev, Tom Kuo, Edward Shuryak and Ismael Zahed. Moreover, 
I am grateful Olaf Kaczmarek, Frithjof Karsch, Pok Man Lo, Krzysztof Redlich, Vladimir Skokov and Chihiro Sasaki
for a fruitful collaboration and many rewarding discussions on the topics discussed here. This work was supported in part by EMMI.

\end{document}